\documentclass[aps,prl,twocolumn,showpacs,amsmath,amssymb,superscriptaddress]{revtex4}
\usepackage{graphicx}
\usepackage{epsf}
\usepackage{bm}

\newcommand{\bi}[1]{\mbox{\boldmath$#1$}}

\begin{document}

\title{
Direct Numerical Simulations of Electrophoresis of Charged Colloids
}
\pacs{82.70.Dd 47.65.-d 82.20.Wt 82.45.-h}

\author{Kang Kim}
\affiliation{
Department of Chemical Engineering, Kyoto University, Kyoto 615-8510, Japan
}
\affiliation{
PRESTO, Japan Science and Technology Agency, 
Kawaguchi 332-0012, Japan
}

\author{Yasuya Nakayama}
\affiliation{
Department of Chemical Engineering, Kyushu University, Fukuoka 819-0395, Japan
}
\affiliation{
PRESTO, Japan Science and Technology Agency, 
Kawaguchi 332-0012, Japan
}

\author{Ryoichi Yamamoto}
\affiliation{
Department of Chemical Engineering, Kyoto University, Kyoto 615-8510, Japan
}
\affiliation{
PRESTO, Japan Science and Technology Agency, 
Kawaguchi 332-0012, Japan
}

\begin{abstract}
We propose a numerical method to simulate electrohydrodynamic phenomena
in charged colloidal dispersions.
This method enables us to compute the time evolutions of
colloidal particles, ions, and host fluids simultaneously 
by solving Newton, advection-diffusion, and Navier--Stokes equations
so that the electrohydrodynamic couplings can be fully taken into account.
The electrophoretic mobilities of charged spherical particles are
calculated in several situations.
The comparisons with approximation theories show quantitative agreements
for dilute dispersions without any empirical parameters, however,
our simulation predicts notable deviations in the case of dense dispersions.
\end{abstract}

\maketitle



Electrohydrodynamic phenomena are of
great importance in physical, chemical, and biological science, and also
in several engineering fields~\cite{Russel}.
In the case of electrophoresis of charged particles for example, 
the particles start to move on the application of external electric fields.
The electric double layer, {\it i.e.} the cloud of counterions 
around charged particles, tends to be deformed and its distribution becomes
anisotropic because of the applied external field and also of the friction 
between ions and fluids.
The electrophoretic mobility of a single colloidal particle is then 
determined by the balance between the electrostatic driving force and 
the hydrodynamic frictional force acting on the particle.
In this situation, the time evolutions of the colloidal particles, 
the ions, and the host fluids are described by 
coupled equations of hydrodynamics (Navier-Stokes) and electrostatics
(Poisson) with proper boundary conditions imposed on the surfaces of the
colloidal particles.
However, the usual numerical techniques of partial differential equations are
hopeless to deal with dynamical evolutions of many-particle systems since
the moving particle-fluid boundary condition must be treated at every
discrete time step.

In late years, efforts to resolve hydrodynamic interactions in
colloidal dispersions attract much attention.
Various advanced methods have been proposed such as the Stokesian
Dynamics (SD)~\cite{JF_Brady1988},
the finite element method (FEM)~\cite{HH_Hu2001},
the Lattice Boltzmann method (LBM)~\cite{AJC_Ladd2001},
the Stochastic Rotation Dynamics~\cite{A_Malevanets1999}, the Fluid Particle
Dynamics (FPD)~\cite{H_Tanaka2000},
and yet another method which treats solid-fluid interaction
efficiently~\cite{T_Kajishima2001}.
Pioneering approaches have been proposed also to simulate 
charged colloidal dispersions without
hydrodynamics~\cite{M_Fushiki1992,H_Lowen1992,H_Lowen1993JCP,J_Chakrabarti2004,J_Dobnikar2003JCP}.
Although extensions have been done to take into account 
the hydrodynamics by using 
SD~\cite{RR_Netz2005}, FEM~\cite{OCTA_MUFFIN}, 
LBM~\cite{J_Horbash2001,R_Wan2003,V_Lobaskin2004JPC,A_Chatterji2005,V_Lobaskin2006},
and FPD~\cite{H_Kodama2004}, quantitatively reliable simulations have
not yet been performed successfully for many particle dispersions
due to the complexity of the system.

Recently, we developed a reliable and efficient numerical method,
called smoothed profile (SP)
method~\cite{Y_Nakayama2005,Y_Nakayama2006},
to resolve the hydrodynamic interactions acting on
solid particles immersed in Newtonian host fluids.
In the SP method,
the original sharp boundaries between colloids and host fluids are 
replaced with diffuse interfaces with finite thickness $\xi$.
This simple modification greatly improve the performance of numerical
computations since it enables us to use the fixed Cartesian grid even
for the problems with moving  boundary conditions.

The SP method is not only applicable to the dispersions in Newtonian
fluids, but particularly suitable for the particle dispersions 
in complex fluids.
It has already been applied successfully to 
liquid crystal colloidal dispersions~\cite{R_Yamamoto2001,R_Yamamoto2004} 
and charged colloidal dispersions~\cite{K_Kim2005}.
Field-particle hybrid simulations were performed, 
where the average direction of the liquid crystal molecules and the
density of ions were treated as coarse-grained continuum
objects while colloids were treated explicitly as particles.
The interaction between fields and particles were taken through the
diffuse interface.
The above methods for the dispersions in complex fluids are, however, 
not yet appropriate for simulating dynamical phenomena since 
hydrodynamic effects are completely neglected.
The purpose of the present study is to establish an efficient and reliable
simulation method applicable for electrohydrodymanic phenomena 
such as electrophoresis by combining our SP methods for 
hydrodynamic~\cite{Y_Nakayama2005,Y_Nakayama2006} 
and electrostatic~\cite{K_Kim2005} interactions.

In the present paper, we briefly outline our numerical modeling
for charged colloidal
dispersions and then demonstrate the reliability of the combined 
SP method by
comparing our numerical results with classical approximation 
theories~\cite{Mvon_Smoluchowski1918,E_Huckel1924,DC_Henry1931,RW_Obrien1978}.
Finally, comparisons are made for the electrophoretic mobilities of 
dense dispersions,
where the simulation results show notable deviations from a mean-field type
theory according to the cell model~\cite{S_Levine1974,H_Ohshima1997}.


Let us consider $N$ spherical particles with radius $a$, the
mass $M_{p}$, and the inertia tensor $\bi{I}_{p}$ in a host fluid
consisting of multi-component ions of species $\alpha$ with
charges $Z_{\alpha}e$, where $e$ is the unit charge.
The local number density of $\alpha$ ion is $C_\alpha(\vec{r},t)$ at a
time $t$.
The total charge on a colloidal particle is $Ze$ and 
distributed uniformly on its surface.
The velocity field of the host fluid is $\vec{v}(\vec{r},t)$.
The position, the translational velocity, and the angular velocity 
of the $i$th particle are $\vec{R_i}$, $\vec{V_i}$, and
$\vec{\Omega}_{i}$, respectively. 
We define the overall profile function
$\phi(\vec{r},t)\equiv\sum_{i=1}^{N}\phi_i(\vec{r},t)$,
where $\phi_{i}\in [0,1]$ is the $i$th particle's profile field which is unity
in the particle domain $|\vec{r}-\vec{R_i}|<a-\xi/2$, zero in the fluid domain
$|\vec{r}-\vec{R_i}|>a+\xi/2$, and have a
continuous diffuse interface within the thin interface domain
$a-\xi/2<|\vec{r}-\vec{R_i}|<a+\xi/2$ whose thickness is $\xi$.
The mathematical definition of $\phi_i$ is given in
Ref.~\cite{Y_Nakayama2005}.
We define the spacial distribution of the surface charge
$e q(\vec{r}) = Ze|\nabla \phi|/4\pi a^2$ using $\phi$, 
then the local density of the total charge is represented smoothly
everywhere in the system by 
$\rho_e(\vec{r}) \equiv \sum_{\alpha}Z_\alpha e C_\alpha + eq$.
The complete dynamics of the system is obtained by solving 
the following time evolution equations~\cite{Y_Nakayama2005,Y_Nakayama2006}.

$i$) The Navier--Stokes equation:
\begin{equation}
\rho(\partial_t
+\vec{v}\cdot\nabla)\vec{v}
=-\nabla p + \eta\nabla^2\vec{v}
-\rho_e\nabla(\Psi+\Psi_{ex})
+\phi\vec{f}_p,
\label{SP_NS}
\end{equation}
with incompressible condition $\nabla\cdot\vec{v}=0$,
where $\rho$ is the mass density, $p$ is the
pressure,
$\eta$ is the shear viscosity of the host fluid,
$\Psi_{ex}=-\vec{E}\cdot\vec{r}$ is the external electric potential due
to the constant electric field $\vec{E}$,
and $\phi \vec{f}_{p}$ represents the body force arising from
the rigidity of the particles~\cite{Y_Nakayama2006}.
The electrostatic potential $\Psi(\vec{r})$ is to be determined 
by solving the Poisson equation
$
\epsilon\nabla^2 \Psi = -\rho_e
$
with the dielectric constant $\epsilon$ of the host fluid.

$ii$) The Newton's and Euler's equations of motions:
\begin{equation}
\dot{\vec{R}}_{i} = \vec{V}_{i},~~
M_{p}\dot{\vec{V}}_{i} =\vec{F}_{i}^{H} + \vec{F}_{i}^{c},~~
\bi{I}_{p}\cdot\dot{\vec{\Omega}}_{i} = \vec{N}_{i}^{H}, 
\end{equation}
where 
$\vec{F}_{i}^{H}$ and $\vec{N}_{i}^{H}$ are the hydrodynamic force
and torque~\cite{Y_Nakayama2006},
and $\vec{F}_{i}^{c}$ is the force arising
from the excluded volume of particles which prevents colloids
from overlapping.
Hereafter, soft-core potential of the truncated Lennard--Jones
potential is adopted for $\vec{F}_{i}^{c}$.
We include the electric driving force due to $\vec{E}$
in $\vec{F}_{i}^{H}$.

$iii$) Advection-diffusion equation:
\begin{equation}
\partial_{t} C^*_\alpha = - \nabla\cdot C^*_\alpha \vec{v}
+f_\alpha^{-1}\nabla\cdot (({\bi I}-\vec{n}\vec{n})\cdot C^*_\alpha\nabla\mu_\alpha),
\label{diffusion_eq}
\end{equation}
where $f_\alpha$ is the ionic friction coefficient,
$\bi{I}$ is the unit tensor, and $\vec{n}$ is a unit-vector field defined by 
$\vec{n}=-\nabla\phi/|\nabla\phi|$.
In our method, the actual density fields of ions are
defined as 
$C_\alpha(\vec{r},t)=(1-\phi(\vec{r},t))C^*_\alpha(\vec{r},t)$
using the auxiliary density field $C^*_\alpha(\vec{r},t)$.
This definition avoids penetration of ions into colloids
explicitly without using artificial potentials, which requires
smaller time increments.
The operator $({\bi I}-\vec{n}\vec{n})$ in Eq.(\ref{diffusion_eq}) 
ensures the conservation of $C_\alpha$ 
since the no-penetration condition,
$\vec{n}\cdot\nabla\mu_\alpha=0$ is directly assigned at the diffuse interface.
Then the charge neutrality  $\int \rho_e d\vec{r}=0$ 
of the total system is guaranteed automatically.
Based on the density functional theory~\cite{JP_Hansen2000,Barrat},
the chemical potential is defined as
\begin{equation}
\mu_\alpha= k_BT\ln C^*_\alpha+Z_\alpha e(\Psi+\Psi_{ex}),
\label{cp}
\end{equation}
where
$k_B$ is the Boltzmann constant and $T$ is the temperature.
If we set $\vec{v}=0$ in 
Eq.~(\ref{diffusion_eq}), 
the equilibrium ($t\rightarrow \infty$)
ionic density is given by
the Boltzmann Eq.
\begin{equation}
C^*_\alpha=\bar C_\alpha \exp[-Z_{\alpha}e(\Psi+\Psi_{ex})/k_BT].
\label{Boltzmann_eq}
\end{equation}
The bulk salt concentration $\bar C_\alpha$ is 
related to the Debye screening length
$\kappa^{-1} =1/\sqrt{4\pi \lambda_B \sum_{\alpha}
Z_\alpha^2 \bar C_\alpha}$, where $\lambda_B=e^2/4\pi \epsilon k_BT$ 
is the Bjerrum length which is 
approximately 0.72nm in water at $25\ {}^\circ\mathrm{C}$.


Simulations have been performed in a three-dimensional cubic box
with the periodic boundary conditions.
The linear dimension is $L/\Delta=64$, where
$\Delta$ is the lattice spacing chosen as a unit length, which is
taken related to the Bjerrum length as $\Delta=4\pi \lambda_B$.
We use the particle radius $a=5$ and 
the thickness parameter $\xi=2$ throughout 
the present simulations.
The host fluid contains 1:1 electrolyte composed of monovalent
counterions ($\alpha=+$) and coions ($\alpha=-$).
The unit of the energy and the electrostatic potential is $k_BT$ and
$k_BT/e$, respectively. The later corresponds to $25.7\mathrm{mV}$ at
$25\ {}^\circ\mathrm{C}$.
The non-dimensional parameter
$m_\alpha=2\epsilon k_BTf_\alpha/3 \eta e^2$ is set to be $m_+=m_-=0.184$, 
which corresponds to the value of KCl solution at
$25\ {}^\circ\mathrm{C}$.
Our unit of time $\tau=\Delta^2f_{+}/k_BT$ corresponds to 
$0.44 \mu\mathrm{sec}$.


\begin{figure}
\includegraphics[scale=0.45]{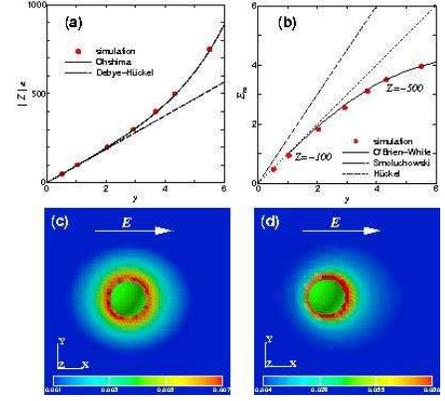}
\caption{
Relationship between surface charge $|Z|e$ and dimensionless zeta potential
$y$ (a). Our numerical data follows nicely on the analytic 
solution of the nonlinear PB equation~\cite{H_Ohshima1982} 
but deviates notably from the 
Debye--H{\"u}ckel linearized theory.
Dimensionless mobility $E_m$ of a single particle is plotted in (b) as a
function of dimensionless zeta potential $y$.
For comparison, results of Smolchowski, Henry, and O'Brien--White for
$\kappa a=0.5$ are plotted.
The color contours in (c) and (d) represent the total ionic charge density
$\sum_{\alpha}eZ_\alpha C_\alpha$ around a single particle
for (c) $Z=-100$ and (d) $Z=-500$.
The electric field is applied in the horizontal ($+x$) direction.
}
\label{zeta}
\end{figure}

We first consider a single charged particle moving with 
the drift velocity $\vec{V}=(-V,0,0)$ 
in a constant electric field $\vec{E}=(E,0,0)$.
The electrophoretic mobility $V/E$ is related to the zeta potential $\zeta$,
which is defined as the electrostatic potential on 
a slipping plane, as
\begin{equation}
V/E = f \epsilon \zeta/\eta
\label{linear_theory}
\end{equation}
when $\zeta$ is small~\cite{Russel}.
The Smolchowski equation $f = 1$ 
is valid in the limiting case 
$\kappa a \to \infty$~\cite{Mvon_Smoluchowski1918},
while the H{\"u}ckel equation $f = 2/3$ 
is valid in the opposite case $\kappa a \to 0$~\cite{E_Huckel1924}.
Henry derived an expression $f = f_H(\kappa a)$ which
is valid for a general value of $\kappa a$~\cite{DC_Henry1931}.
These equations indicate that the mobility is proportional to $\zeta$,
however, this relation tends to fail for larger $\zeta$ where
the relaxation effect due to deformations of electric double layer 
becomes notable.
O'Brien and White proposed an approximation theory which is valid also 
for larger $\zeta$~\cite{RW_Obrien1978}.

We have performed simulations for electrophoresis of a
single particle in linear response regimes $E\lesssim0.15$ 
and compared them with the O'Brien--White theory.
A constant uniform electric field $E=0.1$, which corresponds to 
$2.85\times 10^3 {\rm V/cm}$, was applied.
The terminal $V$ was calculated for $50\leq-Z\leq750$ with $\kappa^{-1}= 10$,
corresponding to the salt concentration $11\mu \mathrm{mol/l}$ at $25\
{}^\circ\mathrm{C}$ in water.
We chose $\nu=\eta/\rho=5$, so the Reynolds number $Re=a V/\nu$ 
remains small.
Both in the O'Brien--White theory and our simulations, 
the zeta potential is assumed to be the electrostatic potential at 
the particle surface, $\zeta=\Psi|_{\rm surface}$.
In our simulations, the surface charges were chosen as
$Z=-50$, $-100$, $-200$, $-300$, $-400$, $-500$, and $-750$,
corresponding to $y=0.525$, $1.044$, $2.035$, $2.927$, $3.692$, $4.332$,
and $5.510$, respectively.
Here the dimensionless zeta potential $y\equiv e\zeta/k_BT$ is introduced.
A relationship between the surface charge $|Z|e$
and the dimensionless zeta potential $y$ is shown in Fig.\ref{zeta}(a), 
where our numerical results are plotted together with an analytic
solution of the nonlinear PB equation~\cite{H_Ohshima1982} 
and the Debye--H{\"u}ckel linearized solution
$\zeta = |Z|e / 4\pi a^2 \epsilon\kappa(1+\kappa a^{-1})$.
We see that our numerical results are consistent with 
the nonlinear PB theory.
In Fig.\ref{zeta}(b), the dimensionless mobility
$E_m\equiv 3e\eta V/2\epsilon k_BT E$ is plotted as a function of
the dimensionless zeta potential with $\kappa a=0.5$.
It is clearly demonstrated that our method reproduced 
the O'Brien--White theory almost perfectly including 
the nonlinear regime $y\geq2$ only within a few percent error.
We emphasize that such a precise agreement with the theory has 
never been obtained by any simulation methods so far proposed.
The distributions of charge density due to counterions and coions 
are shown in Fig.\ref{zeta}(c) for $y=1.044$ and (d) for $y=3.692$.
One can see that the electric double layer is deformed considerably in
the nonlinear regime (d), while it is almost isotropic in the linear
regime (c).
As is mentioned before, the relaxation effect due to the deformed 
double layer causes the nonlinearity in $E_m$.


\begin{figure}
\includegraphics[width=.45\textwidth]{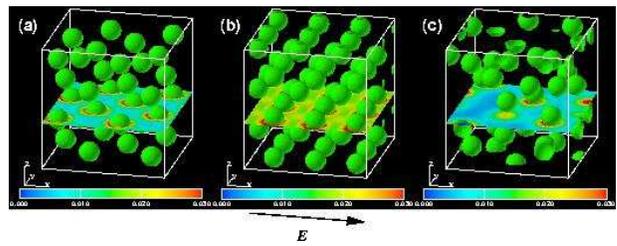}
\caption{
Snapshots of the electrophoresis of dense dispersions 
with (a) FCC, (b) BCC, and (c) random particle configurations.
The color map represents the total ionic charge density 
$\sum_\alpha eZ_\alpha C_\alpha$ in a plane perpendicular
to $z$ axis.
The electric field is applied in $+x$ direction normal
to (1,0,0) face for FCC and BCC.
See movies~\cite{EPAPS}.
}
\label{many_particle}
\end{figure}

Our simulation method is easily applicable to dense dispersions
consisting of many particles.
We thus examined the effect of the particle concentration on 
the electrophoretic mobility.
The linearized theory for a single particle Eq.(\ref{linear_theory})
is still valid for dense dispersions when $E$ is small, however, $f$
is now depending both on $\kappa a$ and $\varphi$.
Simulations were carried out with $Z=-100$ and $E=0.1$ for 
various particle volume fractions $\varphi\equiv 4\pi a^3N/3L^3$
to calculate $f(\kappa a,\varphi)= \eta V/\epsilon\zeta E$.
The Debye length $\kappa^{-1}$ 
is taken to be $5$ and $10$ which correspond to  
$\kappa a=1$ and $0.5$, respectively.
The corresponding salt
concentration is $44\mu\mathrm{mol/l}$ for
$\kappa^{-1}=5$ and $11\mu\mathrm{mol/l}$ for $\kappa^{-1}=10$,
respectively.
Figure~\ref{many_particle} shows typical snapshots 
of the systems with (a) FCC, (b) BCC, and (c) random
configurations~\cite{EPAPS}.
The horizontal color map represents the charge density for $\kappa a=1$
on a cross section perpendicular to $z$ axis.
In the cases of FCC and BCC, $E$ was applied perpendicular to the 
(1,0,0) and (1,1,1) faces, but we obtained very small 
differences only within 1\%.

\begin{figure}
\includegraphics[width=.34\textwidth]{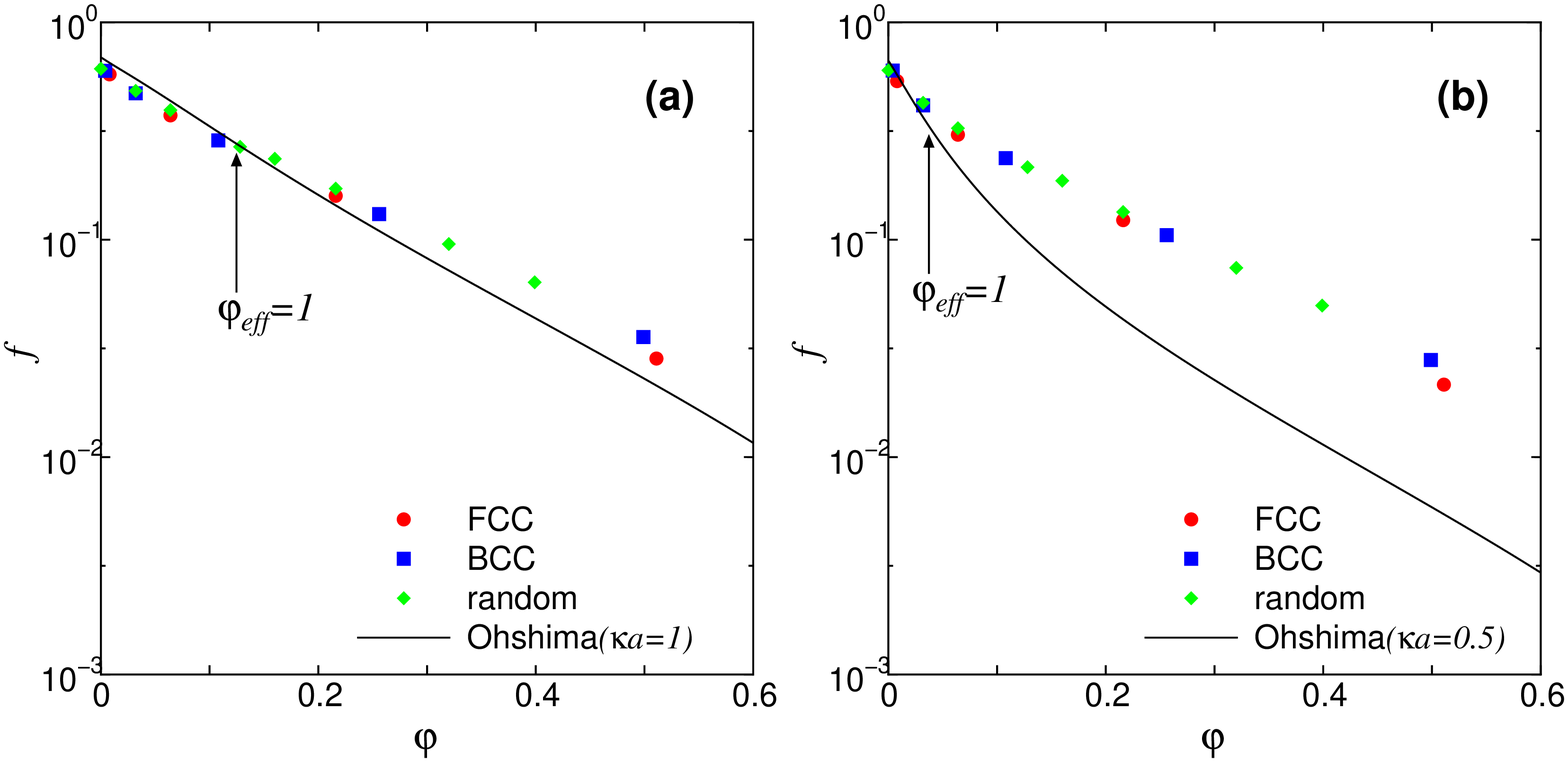}
\caption{
The mobility coefficient $f(\kappa a,\varphi)$ as a
function of the volume fraction $\varphi$ 
in (a) $\kappa a=1$ and (b) $\kappa a=0.5$.
The solid lines represent the approximation theory proposed by 
Ohshima~\cite{H_Ohshima1997}.
The theory is confirmed to be accurate for $\varphi_{eff}\leq1$,
however, tends to deviate from our numerical results 
for $\varphi_{eff}>1$ where 
overlapping of the electric double layers becomes notable. 
}
\label{many_particle2}
\end{figure}

The mobility coefficient $f(\kappa a,\varphi)$
for $\kappa a=1$ and $0.5$ is plotted as a function of 
$\varphi$ in Fig.~\ref{many_particle2} (a) and
(b), respectively.
We found that $f$ decreases rapidly with increasing $\varphi$.
Furthermore, the overall behavior looks almost independent of the particle
configurations.
A theoretical model has been proposed by Levine and Neale 
for the electrophoretic mobility of dense dispersions by using 
the cell model~\cite{S_Levine1974}.
They assumed the situation that a spherical particle 
with radius $a$ is located at the center of a spherical container (cell) 
with radius $b$ and calculated $V$ as a function of 
$\kappa a$ and $\varphi=(a/b)^3$.
Ohshima proposed a simpler expression for the
mobility coefficient $f$ according to the cell model~\cite{H_Ohshima1997}.
Ohshima's prediction is shown in Fig.~\ref{many_particle2} 
(a) and (b) together with our numerical results.
The overall agreement between our simulation and Ohshima's theory 
is better in (a) with a smaller Debye length $\kappa^{-1} =5=a$ than in (b)
with a larger one $\kappa^{-1}=10$.
In both (a) and (b), the simulation results tend to be larger than the
the theory as $\varphi$ increases.
We expect that the deviation is attributable to the occurrence of 
overlapping of the electric double layers for larger $\varphi$
because such an effect is totally neglected in the theory.
To this end, we estimated the effective radius $a+\kappa^{-1}$ 
of the ionically dressed particles and defined the effective volume fraction
$\varphi_{eff}\equiv4\pi (a+\kappa^{-1})^3N/3L^3=(1+(\kappa a)^{-1})^3\varphi$.
As is clearly seen in Fig.\ref{many_particle2} (a) and (b), 
our results agree well with Ohshima' theory for $\varphi_{eff}\leq1$
where the effect of overlapping is small.
However, for $\varphi_{eff}>1$ where 
the overlapping of the electric double layers becomes large, 
deviations between our simulations and the theory become notable.
We emphasize that the present study is the first successful
simulations which provide quantitative data necessary to examine
the reliability of the Ohshima's cell model calculations including their
boundary conditions for electrophoresis in dense colloidal
dispersions.
Our results are consistent with recent studies which also devoted 
to take into account the effects of double layer overlapping~\cite{F_Carrique2003,V_Lobaskin2006,T_Palberg2004}.

In summary, we have developed a unique numerical method for simulating 
electrohydrodynamic phenomena in colloidal dispersions.
The method was first applied to simulate electrophoresis of 
a single spherical particle, and we found that our method can reproduce
the reliable analytical theory proposed 
by O'Brien and White quantitatively.
Simulations were then performed for electrophoresis of colloids in
dense dispersions, and we compared them with the theoretical analysis based on
the cell model.
We found that the cell model is reliable when overlapping of electric
double layers is small but less reliable as overlapping becomes larger.


\begin{thebibliography}{37}
\expandafter\ifx\csname natexlab\endcsname\relax\def\natexlab#1{#1}\fi
\expandafter\ifx\csname bibnamefont\endcsname\relax
  \def\bibnamefont#1{#1}\fi
\expandafter\ifx\csname bibfnamefont\endcsname\relax
  \def\bibfnamefont#1{#1}\fi
\expandafter\ifx\csname citenamefont\endcsname\relax
  \def\citenamefont#1{#1}\fi
\expandafter\ifx\csname url\endcsname\relax
  \def\url#1{\texttt{#1}}\fi
\expandafter\ifx\csname urlprefix\endcsname\relax\def\urlprefix{URL }\fi
\providecommand{\bibinfo}[2]{#2}
\providecommand{\eprint}[2][]{\url{#2}}

\bibitem[{\citenamefont{Russel et~al.}(1989)\citenamefont{Russel, Saville, and
  Schowalter}}]{Russel}
\bibinfo{author}{\bibfnamefont{W.~B.} \bibnamefont{Russel}},
  \bibinfo{author}{\bibfnamefont{D.~A.} \bibnamefont{Saville}},
  \bibnamefont{and} \bibinfo{author}{\bibfnamefont{W.~R.}
  \bibnamefont{Schowalter}}, \emph{\bibinfo{title}{Colloidal Dispersions}}
  (\bibinfo{publisher}{Cambridge University Press, Cambridge},
  \bibinfo{year}{1989}).

\bibitem[{\citenamefont{Brady and Bossis}(1988)}]{JF_Brady1988}
\bibinfo{author}{\bibfnamefont{J.~F.} \bibnamefont{Brady}} \bibnamefont{and}
  \bibinfo{author}{\bibfnamefont{G.}~\bibnamefont{Bossis}},
  \bibinfo{journal}{Ann. Rev. Fluid Mech.} \textbf{\bibinfo{volume}{20}},
  \bibinfo{pages}{111} (\bibinfo{year}{1988}).

\bibitem[{\citenamefont{Hu et~al.}(2001)\citenamefont{Hu, Patankar, and
  Zhu}}]{HH_Hu2001}
\bibinfo{author}{\bibfnamefont{H.~H.} \bibnamefont{Hu}},
  \bibinfo{author}{\bibfnamefont{N.~A.} \bibnamefont{Patankar}},
  \bibnamefont{and} \bibinfo{author}{\bibfnamefont{M.~Y.} \bibnamefont{Zhu}},
  \bibinfo{journal}{J. Comput. Phys.} \textbf{\bibinfo{volume}{169}},
  \bibinfo{pages}{427} (\bibinfo{year}{2001}).

\bibitem[{\citenamefont{Ladd and Verberg}(2001)}]{AJC_Ladd2001}
\bibinfo{author}{\bibfnamefont{A.~J.~C.} \bibnamefont{Ladd}} \bibnamefont{and}
  \bibinfo{author}{\bibfnamefont{R.}~\bibnamefont{Verberg}},
  \bibinfo{journal}{J. Stat. Phys.} \textbf{\bibinfo{volume}{104}},
  \bibinfo{pages}{1191} (\bibinfo{year}{2001}).

\bibitem[{\citenamefont{Malevanets and Kapral}(1999)}]{A_Malevanets1999}
\bibinfo{author}{\bibfnamefont{A.}~\bibnamefont{Malevanets}} \bibnamefont{and}
  \bibinfo{author}{\bibfnamefont{R.}~\bibnamefont{Kapral}},
  \bibinfo{journal}{J. Chem. Phys.} \textbf{\bibinfo{volume}{110}},
  \bibinfo{pages}{8605} (\bibinfo{year}{1999}).

\bibitem[{\citenamefont{Tanaka and Araki}(2000)}]{H_Tanaka2000}
\bibinfo{author}{\bibfnamefont{H.}~\bibnamefont{Tanaka}} \bibnamefont{and}
  \bibinfo{author}{\bibfnamefont{T.}~\bibnamefont{Araki}},
  \bibinfo{journal}{Phys. Rev. Lett.} \textbf{\bibinfo{volume}{85}},
  \bibinfo{pages}{1338} (\bibinfo{year}{2000}).

\bibitem[{\citenamefont{Kajishima et~al.}(2001)\citenamefont{Kajishima,
  Takiguchi, Hamasaki, and Miyake}}]{T_Kajishima2001}
\bibinfo{author}{\bibfnamefont{T.}~\bibnamefont{Kajishima}},
  \bibinfo{author}{\bibfnamefont{S.}~\bibnamefont{Takiguchi}},
  \bibinfo{author}{\bibfnamefont{H.}~\bibnamefont{Hamasaki}}, \bibnamefont{and}
  \bibinfo{author}{\bibfnamefont{Y.}~\bibnamefont{Miyake}},
  \bibinfo{journal}{JSME Int. J. Ser. B} \textbf{\bibinfo{volume}{44}},
  \bibinfo{pages}{526} (\bibinfo{year}{2001}).

\bibitem[{\citenamefont{Fushiki}(1992)}]{M_Fushiki1992}
\bibinfo{author}{\bibfnamefont{M.}~\bibnamefont{Fushiki}}, \bibinfo{journal}{J.
  Chem. Phys.} \textbf{\bibinfo{volume}{97}}, \bibinfo{pages}{6700}
  (\bibinfo{year}{1992}).

\bibitem[{\citenamefont{L{\"o}wen et~al.}(1992)\citenamefont{L{\"o}wen, Madden,
  and Hansen}}]{H_Lowen1992}
\bibinfo{author}{\bibfnamefont{H.}~\bibnamefont{L{\"o}wen}},
  \bibinfo{author}{\bibfnamefont{P.~A.} \bibnamefont{Madden}},
  \bibnamefont{and} \bibinfo{author}{\bibfnamefont{J.-P.}
  \bibnamefont{Hansen}}, \bibinfo{journal}{Phys. Rev. Lett.}
  \textbf{\bibinfo{volume}{68}}, \bibinfo{pages}{1081} (\bibinfo{year}{1992}).

\bibitem[{\citenamefont{L{\"o}wen et~al.}(1993)\citenamefont{L{\"o}wen, Hansen,
  and Madden}}]{H_Lowen1993JCP}
\bibinfo{author}{\bibfnamefont{H.}~\bibnamefont{L{\"o}wen}},
  \bibinfo{author}{\bibfnamefont{J.-P.} \bibnamefont{Hansen}},
  \bibnamefont{and} \bibinfo{author}{\bibfnamefont{P.~A.}
  \bibnamefont{Madden}}, \bibinfo{journal}{J. Chem. Phys.}
  \textbf{\bibinfo{volume}{98}}, \bibinfo{pages}{3275} (\bibinfo{year}{1993}).

\bibitem[{\citenamefont{Chakrabarti et~al.}(2004)\citenamefont{Chakrabarti,
  Dzubiella, and L{\"o}wen}}]{J_Chakrabarti2004}
\bibinfo{author}{\bibfnamefont{J.}~\bibnamefont{Chakrabarti}},
  \bibinfo{author}{\bibfnamefont{J.}~\bibnamefont{Dzubiella}},
  \bibnamefont{and}
  \bibinfo{author}{\bibfnamefont{H.}~\bibnamefont{L{\"o}wen}},
  \bibinfo{journal}{Phys. Rev. E} \textbf{\bibinfo{volume}{70}},
  \bibinfo{pages}{012401} (\bibinfo{year}{2004}).

\bibitem[{\citenamefont{Dobnikar et~al.}(2003)\citenamefont{Dobnikar, Chen,
  Rzehak, and von Gr{\"u}nberg}}]{J_Dobnikar2003JCP}
\bibinfo{author}{\bibfnamefont{J.}~\bibnamefont{Dobnikar}},
  \bibinfo{author}{\bibfnamefont{Y.}~\bibnamefont{Chen}},
  \bibinfo{author}{\bibfnamefont{R.}~\bibnamefont{Rzehak}}, \bibnamefont{and}
  \bibinfo{author}{\bibfnamefont{H.~H.} \bibnamefont{von Gr{\"u}nberg}},
  \bibinfo{journal}{J. Chem. Phys.} \textbf{\bibinfo{volume}{119}},
  \bibinfo{pages}{4971} (\bibinfo{year}{2003}).

\bibitem[{\citenamefont{Kim and Netz}(2005)}]{RR_Netz2005}
\bibinfo{author}{\bibfnamefont{Y.~W.} \bibnamefont{Kim}} \bibnamefont{and}
  \bibinfo{author}{\bibfnamefont{R.~R.} \bibnamefont{Netz}},
  \bibinfo{journal}{Europhys. Lett.} \textbf{\bibinfo{volume}{72}},
  \bibinfo{pages}{837} (\bibinfo{year}{2005}).

\bibitem[{\citenamefont{Yamaue et~al.}(2005)\citenamefont{Yamaue, Sasaki, and
  Taniguchi}}]{OCTA_MUFFIN}
\bibinfo{author}{\bibfnamefont{T.}~\bibnamefont{Yamaue}},
  \bibinfo{author}{\bibfnamefont{M.}~\bibnamefont{Sasaki}}, \bibnamefont{and}
  \bibinfo{author}{\bibfnamefont{T.}~\bibnamefont{Taniguchi}},
  \bibinfo{journal}{Multi-Phase Dynamics Program "Muffin" User's Manual}
  (\bibinfo{year}{2005}), \urlprefix\url{http://octa.jp}.

\bibitem[{\citenamefont{Horbach and Frenkel}(2001)}]{J_Horbash2001}
\bibinfo{author}{\bibfnamefont{J.}~\bibnamefont{Horbach}} \bibnamefont{and}
  \bibinfo{author}{\bibfnamefont{D.}~\bibnamefont{Frenkel}},
  \bibinfo{journal}{Phys. Rev. E} \textbf{\bibinfo{volume}{64}},
  \bibinfo{pages}{061507} (\bibinfo{year}{2001}).

\bibitem[{\citenamefont{Wang et~al.}(2003)\citenamefont{Wang, Fang, Lin, and
  Chen}}]{R_Wan2003}
\bibinfo{author}{\bibfnamefont{R.~Z.} \bibnamefont{Wang}},
  \bibinfo{author}{\bibfnamefont{H.~P.} \bibnamefont{Fang}},
  \bibinfo{author}{\bibfnamefont{Z.}~\bibnamefont{Lin}}, \bibnamefont{and}
  \bibinfo{author}{\bibfnamefont{S.}~\bibnamefont{Chen}},
  \bibinfo{journal}{Phys. Rev. E} \textbf{\bibinfo{volume}{68}},
  \bibinfo{pages}{011401} (\bibinfo{year}{2003}).

\bibitem[{\citenamefont{Lobaskin and D{\"u}nweg}(2004)}]{V_Lobaskin2004JPC}
\bibinfo{author}{\bibfnamefont{V.}~\bibnamefont{Lobaskin}} \bibnamefont{and}
  \bibinfo{author}{\bibfnamefont{B.}~\bibnamefont{D{\"u}nweg}},
  \bibinfo{journal}{J. Phys.: Condens. Matter} \textbf{\bibinfo{volume}{16}},
  \bibinfo{pages}{S4063} (\bibinfo{year}{2004}).

\bibitem[{\citenamefont{Chatterji and Horbach}(2005)}]{A_Chatterji2005}
\bibinfo{author}{\bibfnamefont{A.}~\bibnamefont{Chatterji}} \bibnamefont{and}
  \bibinfo{author}{\bibfnamefont{J.}~\bibnamefont{Horbach}},
  \bibinfo{journal}{J. Chem. Phys} \textbf{\bibinfo{volume}{122}},
  \bibinfo{pages}{184903} (\bibinfo{year}{2005}).

\bibitem[{\citenamefont{Lobaskin et~al.}(2006)\citenamefont{Lobaskin,
  D{\"u}nweg, Medebach, Palberg, and Holm}}]{V_Lobaskin2006}
\bibinfo{author}{\bibfnamefont{V.}~\bibnamefont{Lobaskin}},
  \bibinfo{author}{\bibfnamefont{B.}~\bibnamefont{D{\"u}nweg}},
  \bibinfo{author}{\bibfnamefont{M.}~\bibnamefont{Medebach}},
  \bibinfo{author}{\bibfnamefont{T.}~\bibnamefont{Palberg}}, \bibnamefont{and}
  \bibinfo{author}{\bibfnamefont{C.}~\bibnamefont{Holm}},
  \bibinfo{journal}{cond-mat/0601588}  (\bibinfo{year}{2006}).

\bibitem[{\citenamefont{Kodama et~al.}(2004)\citenamefont{Kodama, Takeshita,
  Araki, and Tanaka}}]{H_Kodama2004}
\bibinfo{author}{\bibfnamefont{H.}~\bibnamefont{Kodama}},
  \bibinfo{author}{\bibfnamefont{K.}~\bibnamefont{Takeshita}},
  \bibinfo{author}{\bibfnamefont{T.}~\bibnamefont{Araki}}, \bibnamefont{and}
  \bibinfo{author}{\bibfnamefont{H.}~\bibnamefont{Tanaka}},
  \bibinfo{journal}{J. Phys.: Condens. Matter} \textbf{\bibinfo{volume}{16}},
  \bibinfo{pages}{L115} (\bibinfo{year}{2004}).

\bibitem[{\citenamefont{Nakayama and Yamamoto}(2005)}]{Y_Nakayama2005}
\bibinfo{author}{\bibfnamefont{Y.}~\bibnamefont{Nakayama}} \bibnamefont{and}
  \bibinfo{author}{\bibfnamefont{R.}~\bibnamefont{Yamamoto}},
  \bibinfo{journal}{Phys. Rev. E} \textbf{\bibinfo{volume}{71}},
  \bibinfo{pages}{036707} (\bibinfo{year}{2005}).

\bibitem[{\citenamefont{Nakayama et~al.}(2006)\citenamefont{Nakayama, Kim, and
  Yamamoto}}]{Y_Nakayama2006}
\bibinfo{author}{\bibfnamefont{Y.}~\bibnamefont{Nakayama}},
  \bibinfo{author}{\bibfnamefont{K.}~\bibnamefont{Kim}}, \bibnamefont{and}
  \bibinfo{author}{\bibfnamefont{R.}~\bibnamefont{Yamamoto}},
  \bibinfo{journal}{cond-mat/0601322}  (\bibinfo{year}{2006}).

\bibitem[{\citenamefont{Yamamoto}(2001)}]{R_Yamamoto2001}
\bibinfo{author}{\bibfnamefont{R.}~\bibnamefont{Yamamoto}},
  \bibinfo{journal}{Phys. Rev. Lett.} \textbf{\bibinfo{volume}{87}},
  \bibinfo{pages}{075502} (\bibinfo{year}{2001}).

\bibitem[{\citenamefont{Yamamoto et~al.}(2004)\citenamefont{Yamamoto, Nakayama,
  and Kim}}]{R_Yamamoto2004}
\bibinfo{author}{\bibfnamefont{R.}~\bibnamefont{Yamamoto}},
  \bibinfo{author}{\bibfnamefont{Y.}~\bibnamefont{Nakayama}}, \bibnamefont{and}
  \bibinfo{author}{\bibfnamefont{K.}~\bibnamefont{Kim}}, \bibinfo{journal}{J.
  Phys.: Condens. Matter} \textbf{\bibinfo{volume}{16}}, \bibinfo{pages}{S1945}
  (\bibinfo{year}{2004}).

\bibitem[{\citenamefont{Kim and Yamamoto}(2005)}]{K_Kim2005}
\bibinfo{author}{\bibfnamefont{K.}~\bibnamefont{Kim}} \bibnamefont{and}
  \bibinfo{author}{\bibfnamefont{R.}~\bibnamefont{Yamamoto}},
  \bibinfo{journal}{Macromol. Theory Simul.} \textbf{\bibinfo{volume}{14}},
  \bibinfo{pages}{278} (\bibinfo{year}{2005}).

\bibitem[{\citenamefont{O'Brien and White}(1978)}]{RW_Obrien1978}
\bibinfo{author}{\bibfnamefont{R.~W.} \bibnamefont{O'Brien}} \bibnamefont{and}
  \bibinfo{author}{\bibfnamefont{L.~R.} \bibnamefont{White}},
  \bibinfo{journal}{J. Chem. Soc. Faraday Trans. 2}
  \textbf{\bibinfo{volume}{74}}, \bibinfo{pages}{1607} (\bibinfo{year}{1978}).

\bibitem[{\citenamefont{von. Smoluchowski}(1918)}]{Mvon_Smoluchowski1918}
\bibinfo{author}{\bibfnamefont{M.}~\bibnamefont{von. Smoluchowski}},
  \bibinfo{journal}{Z. Phys. Chem.} \textbf{\bibinfo{volume}{92}},
  \bibinfo{pages}{129} (\bibinfo{year}{1918}).

\bibitem[{\citenamefont{H{\"u}ckel}(1924)}]{E_Huckel1924}
\bibinfo{author}{\bibfnamefont{E.}~\bibnamefont{H{\"u}ckel}},
  \bibinfo{journal}{Phys. Z.} \textbf{\bibinfo{volume}{25}},
  \bibinfo{pages}{204} (\bibinfo{year}{1924}).

\bibitem[{\citenamefont{Henry}(1931)}]{DC_Henry1931}
\bibinfo{author}{\bibfnamefont{D.~C.} \bibnamefont{Henry}},
  \bibinfo{journal}{Proc. R. Soc. London Ser. A}
  \textbf{\bibinfo{volume}{133}}, \bibinfo{pages}{106} (\bibinfo{year}{1931}).

\bibitem[{\citenamefont{Levine and Neale}(1974)}]{S_Levine1974}
\bibinfo{author}{\bibfnamefont{S.}~\bibnamefont{Levine}} \bibnamefont{and}
  \bibinfo{author}{\bibfnamefont{G.~H.} \bibnamefont{Neale}},
  \bibinfo{journal}{J. Colloid Interface Sci.} \textbf{\bibinfo{volume}{47}},
  \bibinfo{pages}{520} (\bibinfo{year}{1974}).

\bibitem[{\citenamefont{Ohshima}(1997)}]{H_Ohshima1997}
\bibinfo{author}{\bibfnamefont{H.}~\bibnamefont{Ohshima}}, \bibinfo{journal}{J.
  Colloid Interface Sci.} \textbf{\bibinfo{volume}{188}}, \bibinfo{pages}{481}
  (\bibinfo{year}{1997}).

\bibitem[{\citenamefont{Hansen and L{\"o}wen}(2000)}]{JP_Hansen2000}
\bibinfo{author}{\bibfnamefont{J.-P.} \bibnamefont{Hansen}} \bibnamefont{and}
  \bibinfo{author}{\bibfnamefont{H.}~\bibnamefont{L{\"o}wen}},
  \bibinfo{journal}{Annu. Rev. Phys. Chem.} \textbf{\bibinfo{volume}{51}},
  \bibinfo{pages}{209} (\bibinfo{year}{2000}).

\bibitem[{\citenamefont{Barrat and Hansen}(2003)}]{Barrat}
\bibinfo{author}{\bibfnamefont{J.-L.} \bibnamefont{Barrat}} \bibnamefont{and}
  \bibinfo{author}{\bibfnamefont{J.-P.} \bibnamefont{Hansen}},
  \emph{\bibinfo{title}{Basic Concepts for Simple and Complex Liquids}}
  (\bibinfo{publisher}{Cambridge University Press, Cambridge},
  \bibinfo{year}{2003}).

\bibitem[{\citenamefont{Ohshima et~al.}(1982)\citenamefont{Ohshima, Healy, and
  White}}]{H_Ohshima1982}
\bibinfo{author}{\bibfnamefont{H.}~\bibnamefont{Ohshima}},
  \bibinfo{author}{\bibfnamefont{T.~W.} \bibnamefont{Healy}}, \bibnamefont{and}
  \bibinfo{author}{\bibfnamefont{L.~R.} \bibnamefont{White}},
  \bibinfo{journal}{J. Colloid Interface Sci.} \textbf{\bibinfo{volume}{90}},
  \bibinfo{pages}{17} (\bibinfo{year}{1982}).

\bibitem[{EPA()}]{EPAPS}
\bibinfo{howpublished}{See EPAPS Document No. [number will be inserted by
  publisher] for movies of FCC (Fig2a.mpg), BCC (Fig2b.mpg), and random
  (Fig2c.mpg) configurations. For more information on EPAPS, see
  http://www.aip.org/pubservs/epaps.html.}

\bibitem[{\citenamefont{Carrique et~al.}(2003)\citenamefont{Carrique, Arroyo,
  Jim{\'e}nez, and Delgado}}]{F_Carrique2003}
\bibinfo{author}{\bibfnamefont{F.}~\bibnamefont{Carrique}},
  \bibinfo{author}{\bibfnamefont{F.~J.} \bibnamefont{Arroyo}},
  \bibinfo{author}{\bibfnamefont{M.~L.} \bibnamefont{Jim{\'e}nez}},
  \bibnamefont{and} \bibinfo{author}{\bibfnamefont{{\'A}.~V.}
  \bibnamefont{Delgado}}, \bibinfo{journal}{J. Phys. Chem. B}
  \textbf{\bibinfo{volume}{107}}, \bibinfo{pages}{3199} (\bibinfo{year}{2003}).

\bibitem[{\citenamefont{Palberg et~al.}(2004)\citenamefont{Palberg, Medebach,
  Garbow, Evers, Fontecha, Reiber, and Bartsch}}]{T_Palberg2004}
\bibinfo{author}{\bibfnamefont{T.}~\bibnamefont{Palberg}},
  \bibinfo{author}{\bibfnamefont{M.}~\bibnamefont{Medebach}},
  \bibinfo{author}{\bibfnamefont{N.}~\bibnamefont{Garbow}},
  \bibinfo{author}{\bibfnamefont{M.}~\bibnamefont{Evers}},
  \bibinfo{author}{\bibfnamefont{A.~B.} \bibnamefont{Fontecha}},
  \bibinfo{author}{\bibfnamefont{H.}~\bibnamefont{Reiber}}, \bibnamefont{and}
  \bibinfo{author}{\bibfnamefont{E.}~\bibnamefont{Bartsch}},
  \bibinfo{journal}{J. Phys.: Condens. Matter} \textbf{\bibinfo{volume}{16}},
  \bibinfo{pages}{S4039} (\bibinfo{year}{2004}).

\end{thebibliography}

\end{document}